\documentclass[12pt]{iopart}
\usepackage{graphicx,latexsym,stmaryrd,iopams}
 
\begin{document}

\title{Ground state of 1D bosons with delta interaction: link to the BCS model}

\author{M.T. Batchelor\dag \footnote[3]{Corresponding author (murrayb@maths.anu.edu.au) },
X.W. Guan\dag\ and  J.B. McGuire\ddag}
\address{\dag\
{\small Department of Theoretical Physics, RSPSE, and}\\
{\small Centre for Mathematics and its Applications, MSI}\\
{\small The Australian National University, Canberra ACT 0200, Australia}}
\address{\ddag\ 
{\small Physics Department, Florida Atlantic University, Boca Raton, Florida 33431}}


\begin{abstract}
The Bethe roots describing the ground state energy of the
integrable 1D model of interacting bosons with weakly
repulsive two-body delta interactions are seen to satisfy the set of Richardson equations
appearing in the strong coupling limit of an integrable BCS pairing model.
The BCS model describes boson-boson interactions with zero
centre of mass momentum of pairs.
It follows that the Bethe roots of the weakly interacting boson model are given by the 
zeros of Laguerre polynomials. 
The ground state energy and the lowest excitation are obtained explicitly via the Bethe roots. 
A direct link has thus been established, in the context of integrable 1D models,
between bosons interacting via weakly repulsive two-body delta-interactions and 
strongly interacting  Cooper pairs of bosons.
\end{abstract}


\maketitle

There has been a revival of interest in the exactly solved 1D 
model of  interacting bosons \cite{LL}.  
The conventional description of the model is $N$ interacting bosons 
governed by the Hamiltonan
\begin{equation}
{\cal H}=-\sum_{i = 1}^{N}\frac{\partial ^2}{\partial x_i^2}+2\,c \sum_{1\leq i<j\leq N}
\delta (x_i-x_j)
\label{Ham}
\end{equation}
and constrained by periodic boundary conditions to a line of length $L$.
Here $2c$ is the strength of the two-body delta interaction, 
with  $\hbar =2m=1$.
The wave functions are given  in terms of the Bethe ansatz by 
\begin{equation}
\psi(x_1,\cdots, x_N)=\sum_{p}A(p) \exp(\mathrm{i}\sum_jk_{p_j}x_j)
\end{equation} 
in the region $x_1<x_2<x_3<\cdots <x_N$.
The summation extends over all permutations $p$ of momenta 
$\left\{ k_j\right\}$ and $A(p)$ are coefficients depending on $p$. 
The eigenvalues are given by ${\cal E}=\sum_{j=1}^Nk_{j}^2$ where 
$\left\{ k_j\right\}$ satisfy the Bethe equations
\begin{equation}
\exp(\mathrm{i}k_jL)=- \prod^N_{\ell = 1} 
\frac{k_j-k_\ell+\mathrm{i}\, c}{k_j-k_\ell-\mathrm{i}\, c} \qquad \mbox{for} \quad j = 1,\ldots, N.
\label{BE}
\end{equation}
The Bethe roots $k_j$ are known to be real for repulsive interactions, $c > 0$, for which
the ground state energy and excitation spectrum have been studied 
extensively.\footnote{See, e.g., Refs \cite{LL,L, MC, YY, Gaudin, MAT, KOR} and references therein.}

The revival of interest in this model has been inspired by recent experimental and 
theoretical work on low-dimensional trapped boson gases  at ultracold 
temperatures \cite{EXPboson1,EXPboson2,BEC6, BEC3, BEC4, BEC5, BEC7, BEC8}, 
particularly with regard to
highly elongated traps which give possible realizations of a 1D quantum gas as well as 
Bose-Einstein condensates.
In a remarkable recent experiment, a 1D quantum gas was created 
in an optical lattice in which the interactions between trapped ultracold Rb atoms 
were modified to bring about a continuous passage from the weakly interacting regime
to the strongly correlated Tonks-Girardeau gas (Ref.~\cite{EXPboson1} and refs therein).
The key parameter is essentially the ratio $\gamma$ of interaction to kinetic energy.
In this way the pronounced fermionic behaviour of the Tonks-Girardeau gas was
observed in the strong coupling regime \cite{EXPboson1}.

Here we consider asymptotic solutions to the Bethe equations of the interacting boson 
model in the strong and weak coupling limits.
We see that the momentum density distribution agrees with
the observed profiles of the 1D interacting quantum gas \cite{EXPboson1}.
Moreover, we find a link between the Bethe equations
(\ref{BE}) in the weak coupling regime and the Richardson equations
\cite{BCS1} for the BCS boson pairing model in the strong coupling
regime \cite{BCS2,BCS5,OSDR}.
The BCS model describes boson-boson interactions with zero
centre of mass momentum of pairs.
In this way the Bethe roots for the ground state energy of the weakly interacting Bose gas
are characterized by either the roots of Hermite polynomials or appropriate Laguerre polynomials.
In the other direction, we thus see an explicit connection between strongly 
interacting Cooper pairs described by the BCS boson model and bosons interacting 
through weakly repulsive two-body delta interactions.
We begin by looking at the strong coupling limit.

\vskip 5mm
\noindent
{\em{Strong coupling limit}}
 
In the $Lc=\infty$ limit, for the dilute gas ($L>N$) there is a well known connection to non-interacting fermions,
known as the Tonks-Giradeau gas \cite{KG}. 
The Bethe equations (\ref{BE}) reduce to $\exp(\mathrm{i}k_jL)=(-1)^{N-1}$, with solutions
$2\ell\pi /L$ with $\ell$ integer for odd $N$ and half-odd-integer for even $N$. 
Corrections to order $1/c$ can be readily obtained from the asymptotic solutions of the 
Bethe equations (\ref{BE}).
Define $\lambda = Lc/N$.
The roots are real and symmetric about the origin, with
$\left\{\pm k_{2m-1} ,\,m=1,\cdots, \frac{N}{2}\right\}$ for  even $N$, where 
$k_{\ell} =\frac{\ell\pi}{L} \left(1-\frac{2}{\lambda} \right) $.
For odd $N$, the roots are zero and $\left\{\pm k_{2m}, m=1,\cdots, \frac{N-1}{2}\right\}$.
The ground state energy  follows directly from these asymptotic solutions, with
\begin{equation}
\frac{{\cal E}_0}{N} = \frac{1}{3}(N^2-1)\frac{\pi ^2}{L^2}\left(1-\frac{2N}{Lc}\right)^2
\approx \frac{\pi ^2\rho^2}{3}\left(1-\frac{4}{\lambda}\right)\label{E0},
\end{equation}
where  $\rho=N/L$.
This result coincides with that of the perturbation theory approach (see, e.g. Ref \cite{SEN}).
At zero temperature, the chemical potential is given by
$\mu=\pi^2\rho^2-16\pi ^2\rho^3/3c$.
The two-body correlation function $g_2=4\pi^2\rho_0^2/\gamma ^2$
follows from the ground state energy and the Hellmann-Feynman theorem,
with $g_2$ giving the rates of two-body inelastic processes in
1D trapped boson gases \cite{BEC5}. 
Moreover, we see that the momentum density distribution for the 1D $N$-body
interacting bosons is flat due to the symmetric equal-spacing
distribution in momentum space. 
This qualitatively coincides with the
experimental momentum profile of the Tonks-Girardeau gas in the strong
coupling limit, for example, $\gamma \approx 204.5$ \cite{EXPboson1}.
In this regime, the strong repulsive delta potential prevents the
bosons from occupying the same position. 
Therefore, they are spread out and
extend to a larger region in momentum space than in the case of
weakly interacting bosons.

\vskip 5mm
\noindent
{\em{Weak coupling limit}}

The situation in the weak coupling limit is far more interesting. 
The ground state is an analytic function for $c > 0$. 
Considering the limit $Lc<<1$ and after some tedious case by case
calculations for $N=2,3,\ldots,12$, we find that the Bethe roots satisfy the set
\begin{equation}
k_j=\frac{2\pi d_j}{L}+\frac{2c}{L} \, {\sum_{\ell=1}^{N}}'\frac{1}{k_j-k_\ell}, \quad j=1,\cdots N, 
\label{momenta}
\end{equation}
of nonlinear algebraic equations.
Here the summation excludes $j=\ell$ and $d_j=0,\pm 1,\pm 2, \ldots$
denotes excited states for fixed $N$.
The ground state has zero total momentum, with $d_j=0$ for $j=1,\ldots,N$.
These equations are closely related to Stieltjes problems \cite{SD}. 
The ground state energy per particle 
\begin{equation}
\frac{{\cal E}_0}{N}=\frac{c(N-1)}{L},
\label{enrgy1}
\end{equation}
follows directly from (\ref{momenta}).
The nonlinear algebraic equations (\ref{momenta}) for the ground state
were found by Gaudin \cite{Gaudin}, who showed that the $k_j$ are roots
of Hermite polynomials of degree $N$, namely $H_N(k)=0$.

We are also interested in the lowest excited state, 
which has total momentum $k=\frac{2\pi}{L}$. 
Without loss of generality, we may take $d_1=1,\,d_2=\ldots =d_N=0$. 
The lowest excitation energy per particle
\begin{equation}
\frac{{\cal E}_1}{N}\approx \frac{c(N-1)}{L}+\frac{2\pi }{LN}\left(\frac{2\pi }{L}
+\frac{2c(N-1)}{L}+\frac{c^2(N-2)^2}{L2\pi } \right).
\label{enrgyexcit}
\end{equation}
follows from solving equations (\ref{momenta}).
Correspondingly, the largest Bethe root 
is $k_1=\frac{2\pi}{L}+\frac{2c(N-1)}{L}+\frac{c^2(N-2)^2}{L2\pi}$. 
It is clearly seen that the energy gap vanishes in the thermodynamic limit.  
Some numerical exploration reveals that the excitation energy per particle
(\ref{enrgyexcit}) is very accurate for arbitrary number of bosons. 
Other excitation energies depend on the assignments of $d_j$, with total momentum
$k=\sum_j^N {2d_j\pi}/{L}$.

\begin{figure}[t]
\begin{center}
\includegraphics[width=0.6\linewidth]{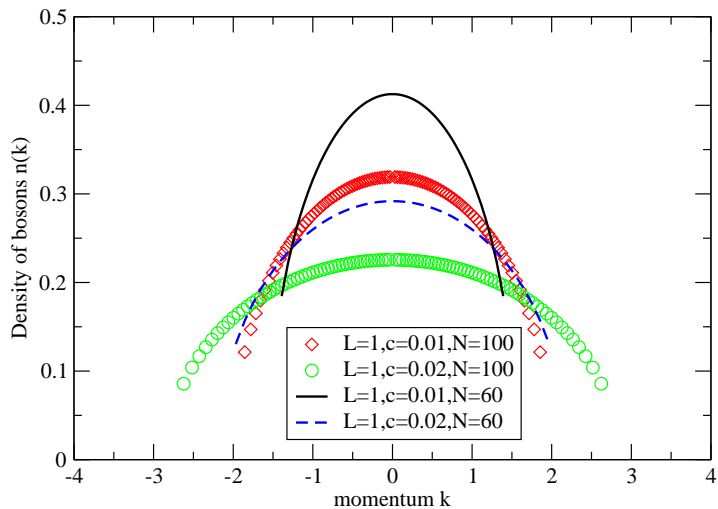}
\end{center}
\caption{The density $n(k)$ of momentum distribution (normalized) for
$N=60$ and $100$ particles: the density distribution function sensitively
depends on interaction strength, particle number and the length $L$
for the 1D boson gas with repulsive delta interaction in the limit $Lc<<1$.}
\label{fig:Boson100}
\end{figure}

The connection between the Bethe roots for the ground state energy and the 
roots of $H_N(k)$ provides a systematic way for studying quantities such as correlations 
and the momentum distribution function.
The normalized momentum density distribution is given by the semi-circle law \cite{Gaudin}
\begin{equation}
n(k)=\frac{L}{2N\pi c}\left(4c\rho-k^2\right)^{\frac{1}{2}}.
\label{density}
\end{equation}
We show the momentum density distribution in \Fref{fig:Boson100} 
obtained by numerically solving the Hermite polynomials of degree $N=60$ and $100$
for different values of $c$.
The results fit very well the analytical expression (\ref{density}). 
A similar semi-circular momentum density distribution has been found in the 
ferromagnetic Heisenberg chain \cite{Dhar}.  
We see that the distribution at small momentum is rather flat, with an almost linearly
decreasing region at large momentum. 
The stronger the interaction strength, the larger the momentum distribution region. 
This reveals a significant signature of the 1D boson gas in the weakly repulsive limit $Lc<<1$. 
Remarkably, this behaviour was recently observed in the
experiment for weakly interacting bosons with $\gamma \approx 0.5$
\cite{EXPboson1}.  
If $c=0$ all the particles condense in the ground state at zero temperture.

In order to link  the 1D boson gas with two-body delta interactions to the 
BCS type of pairing models,  
we consider first an even number of bosons, with $N=2M$.
We also find, to first order in $c$, that the Bethe roots
for the ground state are of the form
\begin{equation}
k_{1,2}=\pm \sqrt{E_1},\,k_{3,4}=\pm \sqrt{E_2},\ldots, \, k_{2M-1,2M}=\pm \sqrt{E_M},\label{P3}
\end{equation}
where the $E_i$ satisfy the equations
\begin{equation}
-\frac{L}{2c} + {\sum_{j=1}^{M}}' \frac{2}{E_i-E_j}=-\frac{1}{2E_i},
\label{bose}
\end{equation}
for $ i = 1, \ldots, M$, where the summation excludes $j=i$.  

Now similar equations have arisen in a number of contexts \cite{BCS2, SD}.
Of particular interest here is the connection between Eq. (\ref{bose}) and Richardson's equations for the 
BCS pairing model \cite{BCS3,BCS4}  in the strong coupling limit \cite{BCS2}. 
However, the precise link between the integrable boson model with weakly repulsive 
delta interaction and the standard BCS model with strong attractive pairing
interaction is quite subtle.
In terms of the Cooper pair energies $E_i$ participating in the scattering process, 
these latter equations are
\begin{equation}
-\frac{1}{\lambda_{\rm BCS} d} + {\sum_{j=1}^{r}}' \frac{2}{E_i-E_j}=\frac{n'}{E_i},
\label{BCS1}
\end{equation}
for  $ i = 1, \ldots, r$.
Here $n'=n-2(m-r)$, where $n$ is the number of unblocked energy levels, 
$m$ is the number of Cooper pairs 
and $r$ is the number of nonvanishing roots to order $\lambda_{\rm BCS}$, 
where $\lambda_{\rm BCS}$ is
the superconducting coupling constant, with $d$ the mean level spacing.

In the BCS model, the parameters $\{E_i\} $ in Eq.~(\ref{BCS1}) represent the   
energies of $r$ Cooper pairs participating in  the scattering. 
Now the ground state of the BCS model in the strong coupling limit has $r=m$ ($n'=n$), 
the first degenerate group of excited states corresponds to $r=m-1$, etc \cite{BCS2}.
Direct comparison between Eqs.~(\ref{bose}) and (\ref{BCS1}) gives $r=M$ 
and $n' = -\frac12$, leading to the ground state where total Cooper pairs is $m=r=M$. 
However, in the strong coupling limit, all of the energy levels collapse into one 
multiply-degenerate level, namely $n=1$.

Obviously, this state appears to lie outside the physical regime of the BCS model \cite{BCS2} 
due to the presence of the negative fractional quantum number $n' = -\frac12$.
The main reason for this is that in the standard BCS
pairing interaction, the electrons are paired into singlets rather
than into triplets with zero centre of mass momentum of pairs.
Therefore the pairing interaction term contains only a pair of
attractive electrons with opposite spin and momenta so that the
degeneracy at each energy state is a doublet, with level degeneracy $\Omega = 2$ \cite{BCS3}.  
In addition, the commutation relation between BCS pairs (hard core bosons) is not the same 
as that for bosons.

Before addressing the correspondence, we can nevertheless make use of the observation \cite{BCS2} 
that the nonvanishing roots of the BCS Eq.~(\ref{BCS1}) are 
given by the zeros of an associated  Laguerre polynomial.
It follows that the Bethe roots of the weakly interacting boson gas in Eq.~(\ref{bose}) 
are given in terms of the associated  Laguerre polynomial $L_n^k(x)$ by
\begin{equation}
L_M^{-\frac{1}{2}}\left(\frac{L E_i}{2c}\right)=0,
\end{equation}
for $ i = 1, \ldots, M$, where 
\begin{equation}
L_n^k(x)=\sum_{m=0}^n \frac{(n+k)!}{(n-m)! (k+m)! m!} \, x^m .
\end{equation}
Explicitly, for $M=1$ ($N=2$), $E_1=c/L$ and for $M=2$ ($N=4$), $E_1=c(3-\sqrt{6})/L,\,E_2=c(3+\sqrt{6})/L$.

For an odd number of  bosons, $N=2M+1$, we find the ground state Bethe roots to be given by
\begin{equation}
k_1=0,\,k_{2,3}=\pm \sqrt{E_1},\,k_{4,5}=\pm \sqrt{E_2},\ldots, k_{2M,2M+1}=\pm \sqrt{E_M},\label{P4}
\end{equation}
where the ${E_i}$ satisfy the equations
\begin{equation}
{\sum_{j=1}^{M}}' \frac{2}{E_i-E_j}-\frac{L}{2c}=-\frac{3}{2E_i},
\label{bose2}
\end{equation}
for $ i = 1, \ldots, M$.
Comparing again with Eq.~(\ref{BCS1}), $n'=-\frac32$ with again $r=m=M$.
%
%
In this case the Bethe roots are given by
\begin{equation}
L_M^{\frac{1}{2}}\left(\frac{L E_i}{2c}\right)=0,
\end{equation}
for $ i = 1, \ldots, M$.
The first few cases are $E_1=3c/L$ for $M=1$ ($N=3$) with $E_1=3c/L$ and
$E_1=c(5-\sqrt{10})/L,\,E_2=c(5+\sqrt{10})/L$ for $M=2$ ($N=5$).

For both the even and odd cases our results indicate that the Bethe roots approach the 
origin proportional to $\sqrt c$ as $c\rightarrow 0$ for fixed $N$ and $L$.
We have checked this directly via numerical solution of the Bethe equations (\ref{BE}).

We now address the precise correspondence between the boson and BCS models.
In view of  the ground state energy ${\cal E}_0=\sum_{j=1}^{N}k_j^2$ of the boson model, 
we can map two bosons with opposite momenta $\pm k_j$ onto one Cooper pair of bosons 
with energy $E_j=2k_j^2$. 
In such a way, we can show that the ground state of 
the boson model with weakly repulsive delta interactions can be described by
the BCS boson pairing model with strong pairing interaction.  
In particular, we need to consider the boson pairing model \cite{BCS5,BCS6}
\begin{equation}
{\cal H}=\sum_{\ell}\epsilon_\ell\hat{n}_\ell+2g\sum_{j,j'=1}^{n}A^{\dagger}_jA_{j'},
\label{Ham-BCS}
\end{equation}
based on the $su(1,1)$ algebra. 
Here $\hat{n}_\ell=\frac{1}{2} a^{\dagger}_\ell a_\ell+\frac{1}{4}$ denotes the particle number
operator at level $\ell$ and $A^{\dagger}_j=\frac{1}{2}a^{\dagger}_ja^{\dagger}_{j}$
creates a boson pair with zero angular momenta at the level $j$.  
The single-particle operators $a^{\dagger}$ and $a$ satisfy Bose
commutation relations.  
Essentially, the single-particle state can be its own time reversal state for bosons at same levels. 
Therefore at every energy level there is a single pair with degeneracy $\Omega=1$ \cite{BCS5}. 
For the boson pairing model, the pair energies are real.
In the above equation, $n$ is the number of unblocked energy levels and
$g$ is the pairing interaction. 
The energy of the pairing model (\ref{Ham-BCS}) is given by 
\begin{equation}
{\cal E}_{{\rm BCS}}=\sum_{\ell}\epsilon_\ell \nu_\ell+\sum_{j=1}^mE_j,
\end{equation}
where the pair energies $E_j$ satisfy Richardson equations of the form \cite{BCS5}
\begin{equation}
-\frac{1}{2g}-\sum_{\ell=1}^n \frac{2d_\ell}{2\epsilon_\ell-E_k}+{\sum_{i=1}^m}'\frac{2}{E_k-E_i}=0.
\label{Bethe}
\end{equation}
Here $m$ denotes the total number of boson pairs and
$d_\ell=\frac{\nu_\ell}{2}+\frac{\Omega_\ell}{4}$ is the effective pair
degeneracy of single-particle level $\ell$, where $\nu_\ell$ denotes the number of
unpaired particles in level $\ell$. 
Linking to the boson pairing
model \cite{BCS5}, here the degeneracy $\Omega_\ell =1$ and $\nu_\ell=0$.

In terms of the Cooper pair energies $E_i$ participating in the scattering process, 
the equations (\ref{Bethe}) are
\begin{equation}
-\frac{1}{2g} + {\sum_{j=1}^{r}}' \frac{2}{E_i-E_j}=-\frac{n'}{E_i},
\label{BCS2}
\end{equation}
for  $ i = 1, \ldots, r$.
Here $n'=\frac{n}{2}+2(m-r)$, where $n$ is the number of unblocked energy levels, 
$m$ is the number of Cooper pairs 
and $r$ is the number of nonvanishing roots to order $g$.
Direct comparison between Eqs.~(\ref{bose}) and (\ref{BCS2}) gives $g=\frac{c}{L}$,
$r=m=M$ and $n = 1$, leading to a ground state of the BCS boson pairing model
(\ref{Ham-BCS}) with  energy
\begin{equation}
{\cal E}_{{\rm BCS}}=\sum_{i=1}^{r}E_i=2gr({n}'+r-1).
\end{equation}
In the strong coupling limit, all energy levels can collapse into one
multiply-degenerate level. It is thus reasonable that $n=1$. 
Indeed, this state appears to lie in the physical regime of the boson
pairing model (\ref{Ham-BCS}). 
The ground state of the weakly interacting boson model thus corresponds to the ground
state of the strongly coupled BCS boson pairing model.
From equations (\ref{bose}) and (\ref{bose2}), the
ground state energy per particle of the weakly interacting Bose gas follows as 
\begin{equation}
\frac{{\cal E}_0}{N}=\sum_{i=1}^{N}\frac{k_i^2}{N}=\frac{2{\cal E}_{{\rm BCS}}}{N}=\frac{c(N-1)}{L}.
\end{equation}
This agrees with Eq.~(\ref{enrgy1}) and the results of \cite{LL} (see also Refs.~\cite{TAK,Wadati}).
We see that it also coincides with the ground state energy per particle 
of the strongly coupled BCS boson pairing model (\ref{Ham-BCS}).

Further, if the Cooper pair is defined \cite{Llano} as 
$
b_{{\bf k K}} = c_{k_2\downarrow}c_{k_1\uparrow},\,\,\,b_{{\bf k K}}^{\dagger}
=c^{\dagger}_{k_1\uparrow}c^{\dagger}_{k_2\downarrow}
$
where ${\bf k}\equiv \frac{1}{2}(k_1-K_2)$ and ${\bf K}\equiv (k_1+K_2)$, it
is shown that $[b_{{\bf k K}},b_{{\bf {k}' K}}^{\dagger}]=0$ for 
${\bf k}\neq {\bf {k}'}$,
while $ [b_{{\bf k K}},b_{{\bf {k}' K}}]=[b_{{\bf k K}}^{\dagger},b_{{\bf {k}' K}}^{\dagger}]=0$ always hold. 
The single operators $c^{\dagger}$ and $c$ satisfy Fermi anticommutation relations.  
If the relative momentum vector of 
paired electrons lies inside the overlap of the two spherical shells
in momentum space, the Cooper pair with nonzero centre of mass momentum,
i.e. ${\bf K}\equiv (k_1+K_2)$, is possible and physical \cite{Llano}.  
This means that the Cooper pairs can occupy a state of
energy $\epsilon_{{\bf K}}$ with same ${\bf K}$ but different values of ${\bf k}$. 
Thus such Cooper pairs act as bosons and can collapse into a Bose condensate. 
In this BCS pairing model, the degeneracy of single-particle level $\Omega_\ell$ 
would not be restricted to be even such that the quantum number $n'$ in the equation (\ref{BCS1})
can be fractional and negative. 
In this case, the single-particle states for a Cooper pair are not time-reversal eigenstates.  
For the standard BCS model, ${\bf K}=0$, such that the state ${\bf k} \!\! \uparrow$
is occupied so is the state $-{\bf k} \!\! \downarrow$. 
Therefore we can expect that the BCS model of fermion pairs with nonzero centre-mass 
momentum would exhibit some of the behaviour of the 1D boson gas with two-body
delta interactions.

In conclusion we have looked in detail at the many-body solution of the Bethe equations
for the interacting boson model in the strongly and  weakly repulsive coupling regimes.
We found a remarkable connection with the Richardson equations for the strong coupling
limit of the boson pairing model,
suggesting a direct link between strongly interacting Cooper pairs and weakly repulsive bosons.
This link deserves further investigation, particularly in light
of the revival of interest in the interacting Bose gas and the BEC/BCS crossover \cite{cross}.

\vskip 5mm
{\bf Acknowledgments}.  
MTB thanks the Physics Department at FAU, where this work was begun, 
for their kind hospitality. 
In turn JBM acknowledges the hospitality of the CMA and the 
Theoretical Physics Department at ANU.
We  thank G. Sierra, Angela Foerster, Norman  Oelkers,  Rudolf A. R\"{o}mer and Huan-Qiang Zhou  
for some  helpful discussions.
This work has been supported by the Australian Research Council and the
Florida Atlantic University Foundation.

\vskip 10mm


\end{document}